\documentclass{svproc}
\usepackage{url}

\usepackage{epsfig}
\usepackage{subfigure}
\usepackage{amssymb}
\usepackage{amstext}
\usepackage{amsmath}

\begin{document}
\mainmatter              
\title{Machine learning and control engineering: \\  The model-free case}

\titlerunning{Model-free control}

\author{Michel Fliess\inst{1,3} \and  C\'{e}dric Join\inst{2,3}}

\authorrunning{Michel Fliess and C\'{e}dric Join} 

\institute{
{LIX (CNRS, UMR 7161), \'Ecole polytechnique, 91128 Palaiseau, France. \\ \email{ Michel.Fliess@polytechnique.edu }}
\and
{CRAN (CNRS, UMR 7039)), Universit\'{e} de Lorraine, BP 239, \\ 54506 Vand{\oe}uvre-l\`{e}s-Nancy, France. \\
\email{Cedric.Join@univ-lorraine.fr}}\vspace{.2cm}
\and
{AL.I.E.N., 7 rue Maurice Barr\`{e}s, 54330 V\'{e}zelise, France. \\
        \email{ \{michel.fliess, cedric.join\}@alien-sas.com}}
        }

   \maketitle     

%

\begin{abstract}
This paper states that Model-Free Control (MFC), which must not be confused with Model-Free Reinforcement Learning, is a new tool for Machine Learning (ML). MFC is easy to implement and should be substituted in control engineering to ML via Artificial Neural Networks and/or Reinforcement Learning. A laboratory experiment, which was already investigated via today's ML techniques, is reported in order to confirm this viewpoint.
\end{abstract}

%
%

\section{Introduction}
The huge popularity today of \emph{Machine Learning} (\emph{ML)} is due to many beautiful achievements of \emph{Artificial Neural Networks} (\emph{ANNs}) (see, \textit{e.g.}, \cite{lecun1}, \cite{lecun2}, \cite{sej}) and \emph{Reinforcement Learning} (\emph{RL}) (see, \textit{e.g.}, \cite{sutton}). Let us quote \cite{recht}: ``Reinforcement Learning is the subfield of machine learning that studies how to use past data to enhance the future manipulation of a dynamical system. A control engineer might be puzzled by such a definition and interject that this precisely the scope of control theory. That the RL and the control theory communities remain practically disjoint has led to the co-development of vastly different approaches to the same problems. However, it should be impossible for a control engineer not to be impressed by the recent successes of the RL community such as solving Go \cite{go}.'' Many concrete case-studies have already been investigated: see, \textit{e.g.}, \cite{anderson}, \cite{baumeister}, \cite{brunton}, \cite{bucci}, \cite{chen}, \cite{cheon}, \cite{dierks}, \cite{duriez}, \cite{hwang}, \cite{kahn}, \cite{lambert}, \cite{li}, \cite{lucia}, \cite{luo}, \cite{ma}, \cite{miller}, \cite{mnih}, \cite{moe}, \cite{nicol}, \cite{qu}, \cite{rabault}, \cite{radac}, \cite{stalph}, \cite{wang1}, \cite{wang2}, \cite{weislander}, \cite{wu}, \cite{zhang1}, \cite{zhang2}, \cite{zhang3}, \cite{zhang4}, \cite{zhu}. Although those works are most promising, they show that ANNs and RL have perhaps not provided in this field such stunning successes as they did elsewhere.
\begin{remark}
The connection of RL with \emph{optimal control} is known since ever (see, \textit{e.g.}, \cite{bu}, \cite{kiumarsi}, \cite{recht}). According to \cite{matni1}, \cite{matni2} tools stemming from advanced control theory should enhance RL in general.
\end{remark}

This communication suggests another route: \emph{Model-Free Control} (\emph{MFC}) in the sense of \cite{csm}.
\begin{remark}
The meaning of \emph{Model-Free Reinforcement Learning} is quite distinct. In model-free RL there is no transition probability distribution, \textit{i.e.}, no \emph{model} (see, \textit{e.g.}, \cite{sugiyama}, \cite{sutton}). \emph{Q-learning} is an example which has been used several times in control engineering.
\end{remark}
MFC, which is easy to implement both from software \cite{csm} and hardware \cite{hardware} viewpoints, leads to many acknowledged applications (see the bibliographies in \cite{csm}, \cite{bara} for most references until 2018).\footnote{Some applications are patented. Others have been published more recently (see, \textit{e.g}, \cite{abb}, \cite{abou}, \cite{barth1}, \cite{barth2}, \cite{clouatre1}, \cite{clouatre2}, \cite{clouatre3}, \cite{had}, \cite{han}, \cite{hat}, \cite{hong}, \cite{kizir}, \cite{mich}, \cite{nd}, \cite{plumejeau}, \cite{qin}, \cite{rampazzo}, \cite{rocher}, \cite{san}, \cite{tich}, \cite{villagra}, \cite{wang0}, \cite{wang3}, \cite{yang}, \cite{zhang1bis}, \cite{zhang2bis}).} The relationship with ML is sketched below.

Consider a system $S$ with a single input $u$ and a single output $y$. Under rather weak assumptions $S$ may be approximated (see \cite{csm} and Section \ref{mfc}) by the \emph{ultra-local model}:
\begin{equation}
\dot{y}(t) = F(t) + \alpha u(t) \label{1}
\end{equation}
where $F$ encompasses not only the poorly known structure of $S$ but also the disturbances. Since $\alpha$ is a constant that is easy to nail down (see Section \ref{ultra}), the main task is to determine the time-dependent quantity $F(t)$. A real-time estimate  
$F_{\text{est}}(t)$ is given thanks to a new parameter identification technique \cite{sira1}, \cite{sira2}, \cite{sira} by the following integral of the input-output data 
\begin{equation}\label{int}
{\tiny F_{\text{est}}(t)  =-\frac{6}{\tau^3}\int_{t-\tau}^t \left\lbrack (\tau -2\sigma)y(\sigma)+\alpha\sigma(\tau -\sigma)u(\sigma) \right\rbrack d\sigma }
\end{equation}
where $\tau > 0$ is small. Formula \eqref{int}, which ought to be viewed as a kind of \emph{unsupervised learning} (see, \textit{e.g.}, \cite{russel}) procedure, takes into account the time arrow: the structure of $S$ and especially the disturbances might be time-varying in an unexpected way. Moreover the unavoidable corrupting noises are attenuated by the integral, which is a low pass filter (see Remark \ref{noise}). Associate the feedback loop \cite{csm}
\begin{equation}\label{ip}
u = - \frac{F_{\text{est}} - \dot{y}^\star + K_P e}{\alpha}
\end{equation}
where
\begin{itemize}
\item $y^\star$ is a reference trajectory,
\item $e = y - y^\star$ is the tracking error,
\item $K_P \in \mathbb{R}$ is a gain.
\end{itemize}
It has been baptised \emph{intelligent proportional controller} (\emph{iP}), already some time ago \cite{intel}: this unsupervised learning permits not only to track the reference trajectory but also to limit the effects of the disturbances and of the poor system understanding (see Section \ref{propor} for further details).
Note that ANNs and RL, and the corresponding methods from computer sciences and mathematics, are not employed.\footnote{It is perhaps worth mentioning that some other fields of computer sciences might benefit from MFC (see, \textit{e.g.}, \cite{bekcheva,iot}).}

In order to support our viewpoint a lab experiment has been selected. A half-quadrotor is available to one of the authors (C.J.). 
Moreover quadrotors and half-quadrotors have been already examined via ANNs and RL: see, \textit{e.g.}, \cite{hwang}, \cite{lambert}, \cite{radac}, \cite{weislander}.\footnote{Numerous other references do not use any traditional AI techniques.} Our results are not only excellent but also easy to obtain. The interested reader is invited to compare with the above references. It has been moreover shown \cite{clouatre3} that the performances of MFC with respect to quadrotors are  superior to those of PIDs (see, \textit{e.g.}, \cite{astrom}, \cite{murray}).\footnote{\emph{Proportional-Integral-Derivative} (\emph{PID}) controllers ought to be regarded as the ``bread and butter'' of control engineering!} 

This communication is organized as follows.  MFC is reviewed in Section \ref{mfc}. Section \ref{exper} discusses the lab experiment. Some concluding remarks may be found in Section \ref{conclusion}.

\section{MFC as a ML technique}\label{mfc}
\subsection{The input-output system as a functional}
Consider for notational simplicity a
SISO system, \textit{i.e.}, a system
with a single control variable $u(t)$ and a single output variable $y(t)$, where $t \geq 0$ is the time. Even
without knowing any ``good'' mathematical model we may assume that
the system corresponds to a \emph{functional} (see, \textit{e.g.}, \cite{kolmogorov}), \textit{i.e.}, a function of functions,
\begin{equation}\label{functional}
y(t) = \mathcal{F}\left( u({\frak{t}}) ~ | ~ 0 \leq {\frak{t}} \leq t \right)
\end{equation}
$\mathcal{F}$ depends not only on initial conditions at $t = 0$, but also on the unavoidable disturbances.
\subsection{The ultra-local model}\label{ultra}
It has been demonstrated \cite{csm} that, under mild assumptions, the input-output behavior \eqref{functional} may be well approximated by the \emph{ultra-local model}: 
\begin{equation*}
y^{(n)}(t) = F(t) + \alpha u(t) \label{n}
\end{equation*}
where the order $n \geq 1$ of derivation is in all known examples equal to $1$ or $2$. In most concrete case-studies, $n = 1$. The case $n = 2$ arises, for instance, with weak frictions \cite{csm} (see, \textit{e.g.}, \cite{menhour} for a concrete case-study). Consider from now on only the case $n = 1$, \textit{i.e}, Equation \eqref{1}, which works well in Section \ref{exper}:
\begin{itemize}
\item  The time-dependent quantity $F$ is not only encompassing the internal structure of the system, which may be poorly known, but also the disturbances. Write $F_{\text{est}}$ its estimate which is derived in Section \ref{F}.
\item The constant $\alpha \in \mathbb{R}$ is chosen by the practitioner such that the three terms in Equation \eqref{1} are of the same magnitude. A precise determination of $\alpha$ is therefore meaningless.  
In practice $\alpha$ is easily chosen via two possible approaches: 
\begin{itemize}
\item the absolute value $|\frac{\alpha u(t)}{y(t)}|$ is not too far from $1$,
\item trial and error, \textit{i.e.}, a kind of \emph{supervised learning} (see, \textit{e.g.}, \cite{russel}).
\end{itemize}
\end{itemize}

\subsection{Intelligent proportional controllers}\label{propor}
Close the loop in Equation \eqref{1} with the iP \eqref{ip}.
Equations \eqref{1} and \eqref{ip} yield
\begin{equation*}
	\dot{e} + K_P e = F - F_{\text{est}}
			\label{equa15}
	\end{equation*}
If the estimation $F_{\text{est}}$ is ``good'': $F - F_{\text{est}}$ is ``small'', \textit{i.e.}, $F - F_{\text{est}} \simeq 0$,  then $\lim_{t \to +\infty} e(t) \simeq 0$ if $K_P > 0$. It implies that the tuning of $K_P$ is quite straightforward.\footnote{See, \textit{e.g.}, in \cite{villagra} an optimization procedure, which, in some sense, is closer to today's viewpoint on ML.} This is a major benefit when
compared to the tuning of ``classic'' PIDs (see, \textit{e.g.},
\cite{astrom}, \cite{murray}).

\subsection{ML via the estimation of $F$}\label{F}
Any function, for instance $F$ in Equation \eqref{1}, may be approximated under a weak integrability assumption by a piecewise constant function (see, \textit{e.g.}, \cite{bourbaki}). The estimation techniques below are borrowed from \cite{sira1}, \cite{sira2}, \cite{sira}.
\subsubsection{First  approach}
Rewrite Equation \eqref{1}  in the operational domain (see, \textit{e.g.}, \cite{yosida}): 

\begin{equation*}
sY = \frac{\Phi}{s}+\alpha U +y(0)
			\label{equa16}
\end{equation*}
where $\Phi$ is a constant. We get rid of the initial condition $y(0)$ by multiplying both sides on the left by $\frac{d}{ds}$:
\begin{equation*}
Y + s\frac{dY}{ds}=-\frac{\Phi}{s^2}+\alpha \frac{dU}{ds}
			\label{equa17}
	\end{equation*}
Noise attenuation is achieved by multiplying both sides on the left by $s^{-2}$. It yields in the time domain the real-time estimate Formula \eqref{int} thanks to the equivalence between $\frac{d}{ds}$ and the multiplication by $-t$, where $\tau > 0$ might be quite small. This integral, which is a low pass filter, may of course be replaced in practice by a classic digital linear filter.

\subsubsection{Second approach}\label{2e}
Close the loop with the iP \eqref{ip}. It yields:

\begin{equation*}
F_{\text{est}}(t) = \frac{1}{\tau}\left[\int_{t - \tau}^{t}\left(\dot{y}^{\star}-\alpha u
- K_P e \right) d\sigma \right] 
			\label{equa18}
	\end{equation*}
\begin{remark}\label{noise}
\emph{Noises}, which are usually described in engineering and, more generally, in applied sciences via probabilistic and statistical tools, are related here to quick fluctuations around $0$ \cite{bruit}: The integral of any such noise over a finite time interval is close to $0$. The robustness with respect to corrupting noises is thus explained. See, \textit{e.g.}, \cite{beltran}, \cite{sira} for concrete applications in signal processing where the parameter estimation techniques of \cite{sira1}, \cite{sira2}, \cite{sira} have been employed.
\end{remark}

\subsection{MIMO systems}\label{mimo}
Consider a multi-input multi-output (MIMO) system with $m$ control variables $u_i$ and $m$ output variables $y_i$, $i = 1, \dots, m$, $m \geq 2$. It has been observed in \cite{toulon} and confirmed by all encountered concrete case-studies (see, \textit{e.g.}, \cite{wang0}), that such a system may usually be regulated via $m$ monovariable ultra-local models:
\begin{equation}\label{multi}
y_{i}^{(n_i)} = F_i + \alpha_i u_i
\end{equation}
where $F_i$ may also depend on $u_j$, $y_j$, and their derivatives, $j \neq i$.

\section{Experiments: A half-quadrotor}\label{exper}

\subsection{Process description}
\begin{figure*}[!ht]
\centering%
\includegraphics[width=0.90\textwidth]{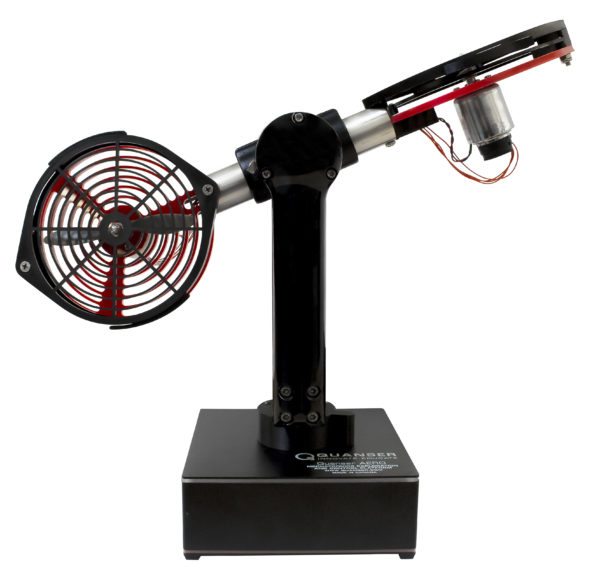}
\caption{The Quanser AERO half-quadrotor}\label{maquette}
\end{figure*}

Our half-quadrotor (see Fig. \ref{maquette}), called AERO, is manufactured by Quanser.\footnote{See the link {\tt https://www.quanser.com/products/quanser-aero/}} Two motors driving the propellers, which might turn clockwise or not, are controlling the angular positions: the azimuth (horizontal) velocity and pitch (vertical) position of the arms.  Outputs $y_1$ and $y_2$ ares respectively mesures of the azimuth velocity (rad/ms) and pitch position (rad). Write $v_i$, $i = 1, 2$, the supply voltage of motor $i$, where $- 24{\rm v} \leq v_i \leq 24{\rm v}$ (volt).  Measures and control inputs are updated each $10$ms. 

\subsection{Control}
It is clear that $y_i$, $i = 1, 2$, is mainly influenced by $v_i$. Equations \eqref{multi} and \eqref{ip} become
\begin{equation}
\label{quad}
\dot{y}_{i}= F_i + \alpha_i u_i
\end{equation}
\begin{equation}\label{quadip}
u_i = - \frac{F_{i, \text{est}} - \dot{y}^\star + K_{P, i} e_i}{\alpha_i}
\end{equation}
The control variable $u_i$ in Equation \eqref{quad} is defined by 
\begin{itemize}
\item if $u_i > 0$, then $v_i=10+u_i$, 
\item if $u_i  <  0$, then $v_i=-10+u_i$.
\end{itemize}
In Equations \eqref{quad}-\eqref{quadip}, set $\alpha = 0.001$, $K_{P, 1} = 0.5$, $\alpha_2 =  5$, $K_{P, 2} = 500$. Everything is programed in $C\#$ and stored in the server.


\subsection{Experiments}
\subsubsection{Nominal half-quadrotor} 
\begin{figure*}[!b]
\centering%
\subfigure[\footnotesize Azimuth velocity (blue $--$), reference trajectory (red $- -$)]
{\epsfig{figure=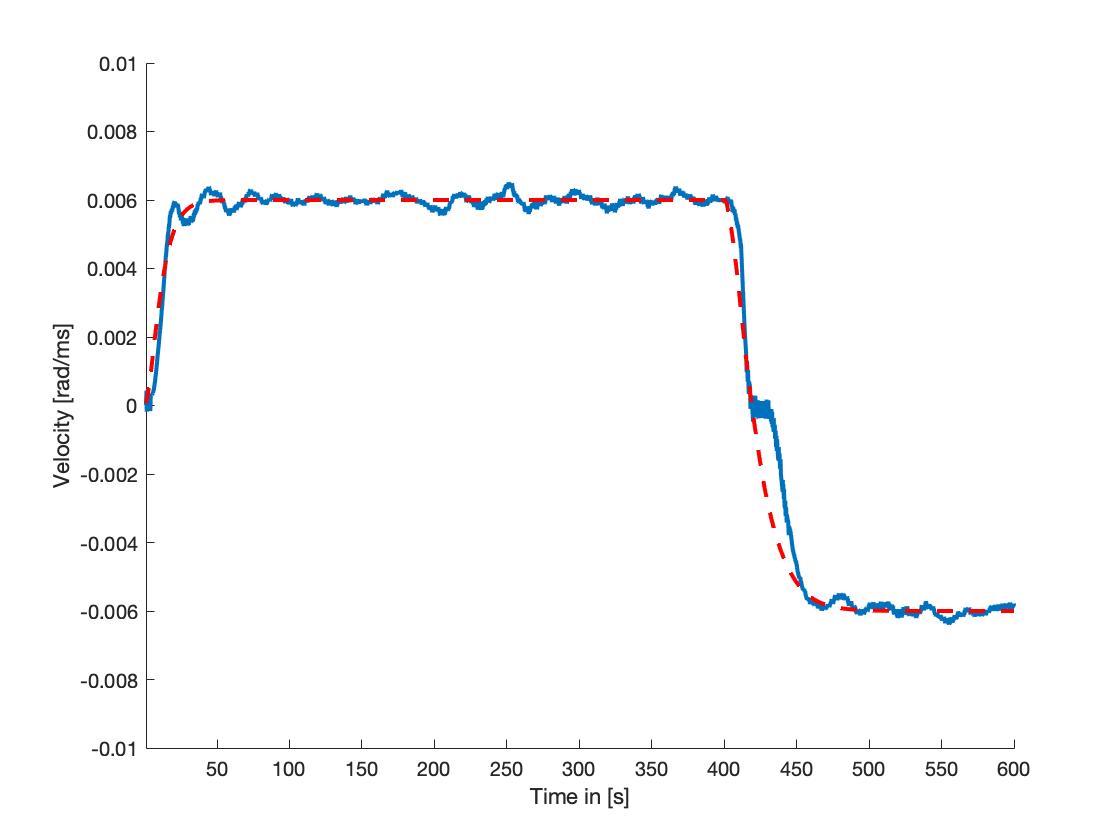,width=1\textwidth}}
\subfigure[\footnotesize Control $u_1$]
{\epsfig{figure=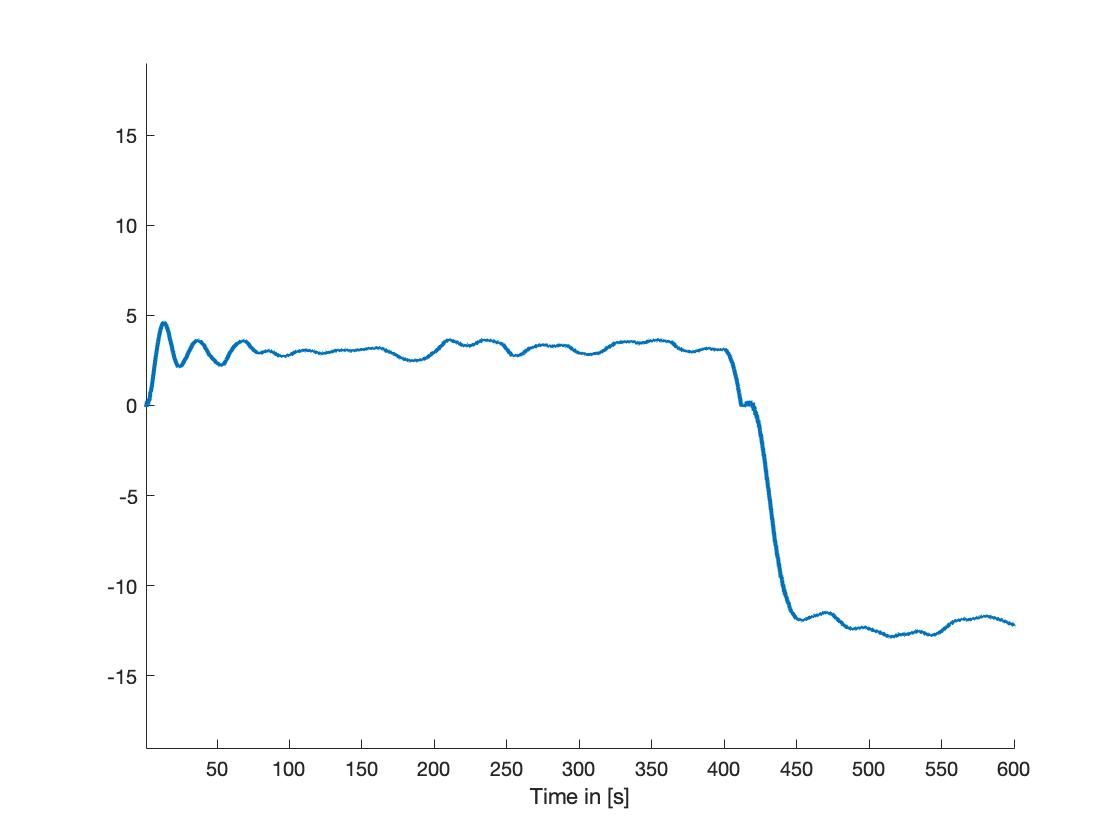,width=1\textwidth}}
\caption{Scenario 1: Azimuth}\label{S11}
\end{figure*}

\begin{figure*}[!b]
\centering%
\subfigure[\footnotesize Pitch position (blue $--$), reference trajectory (red $- -$) ]
{\epsfig{figure=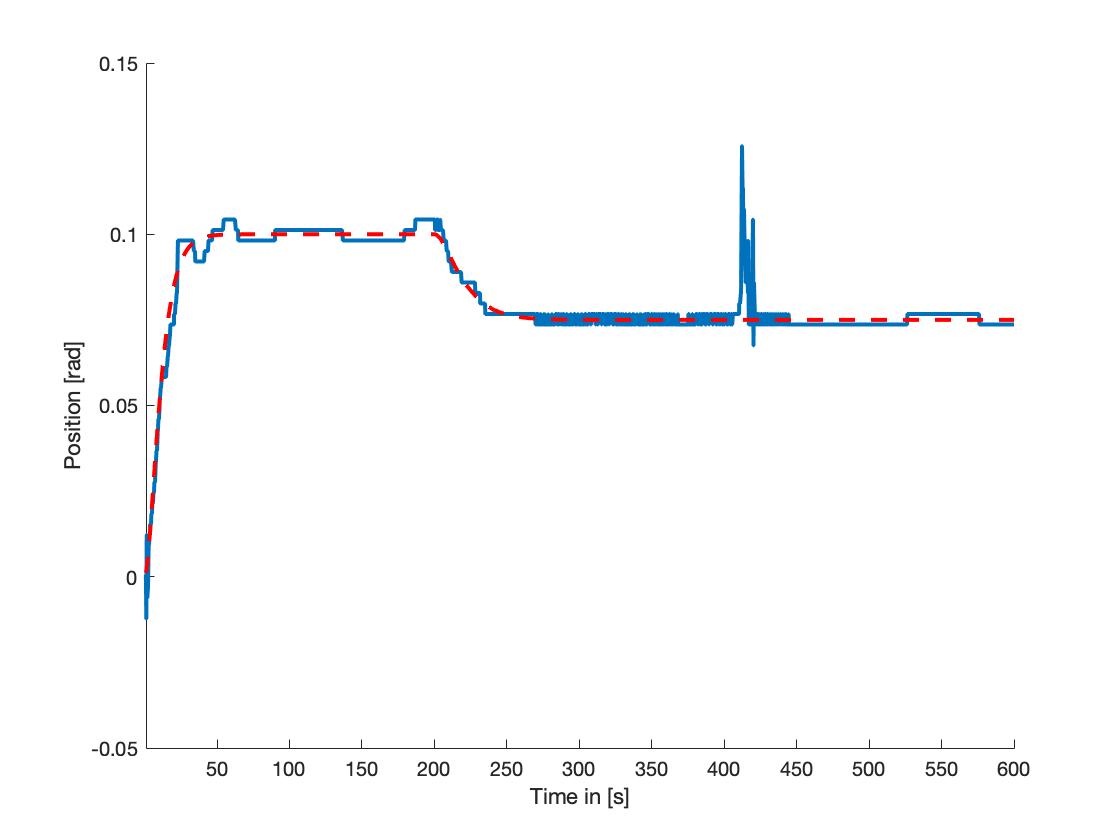,width=1\textwidth}}\\
\subfigure[\footnotesize Control $u_2$]
{\epsfig{figure=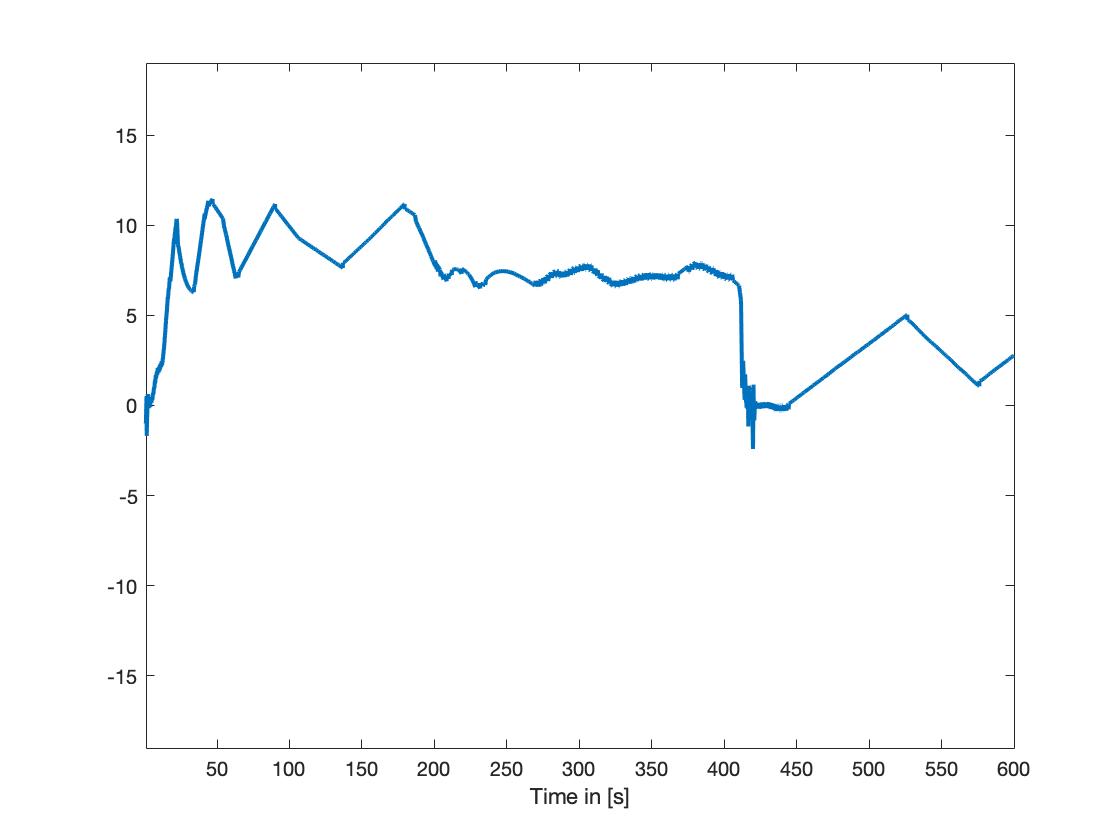,width=1\textwidth}}
\caption{Scenario 1: Pitch}\label{S12}
\end{figure*}

\begin{figure*}[!b]
\centering%
\subfigure[\footnotesize Azimuth velocity (blue $--$), reference trajectory (red $- -$)]
{\epsfig{figure=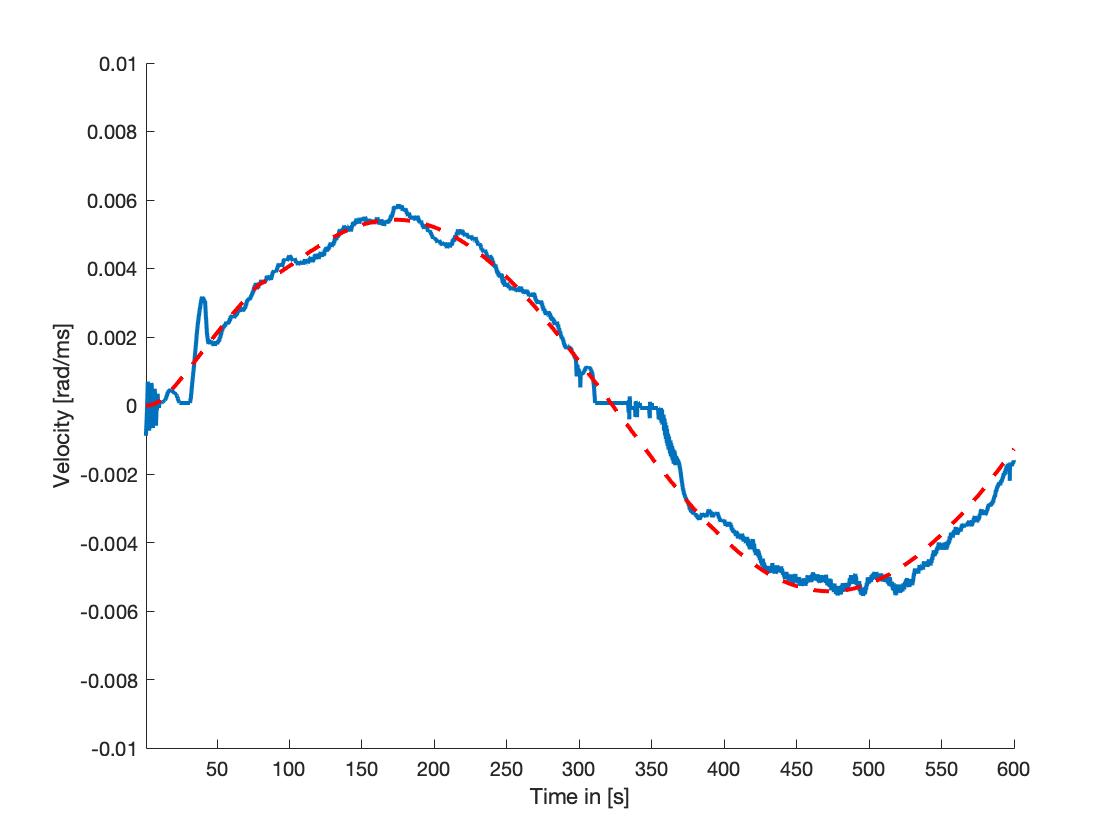,width=1\textwidth}}
\subfigure[\footnotesize Control $u_1$]
{\epsfig{figure=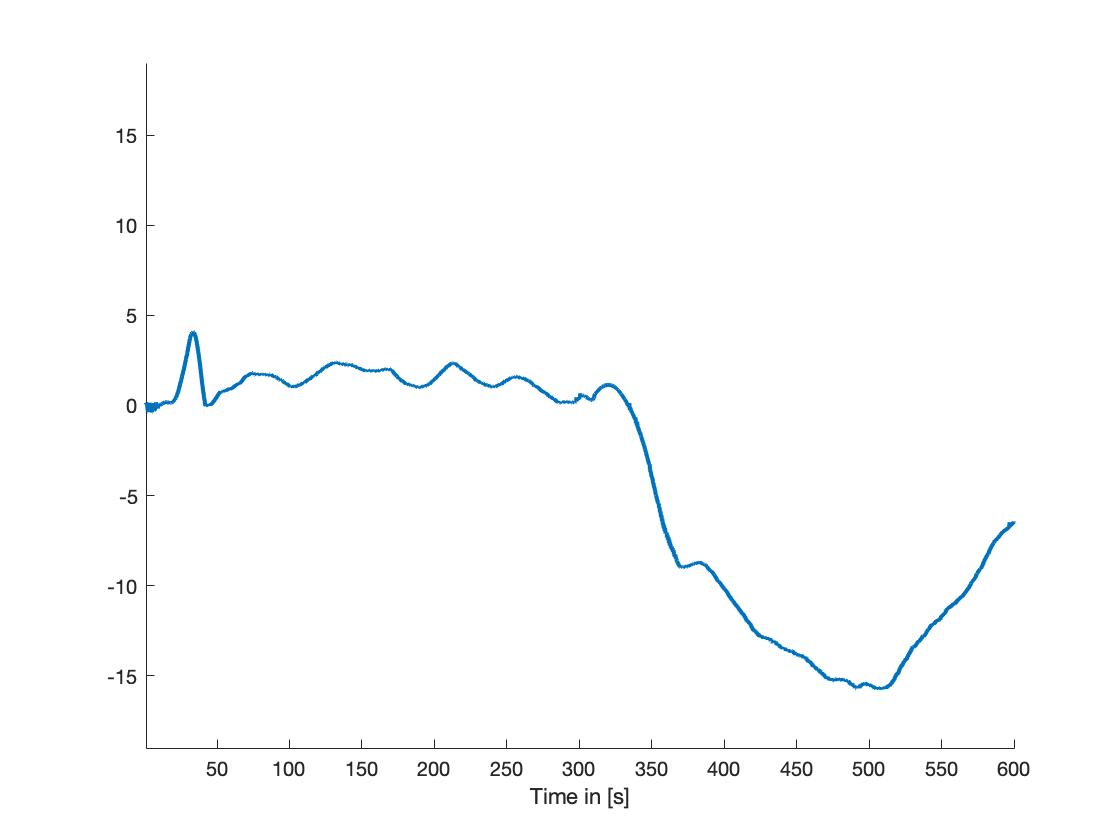,width=1\textwidth}}
\caption{Scenario 2: Azimuth}\label{S21}
\end{figure*}

\begin{figure*}[!b]
\centering%
\subfigure[\footnotesize Pitch position (blue $--$), reference trajectory (red $- -$) ]
{\epsfig{figure=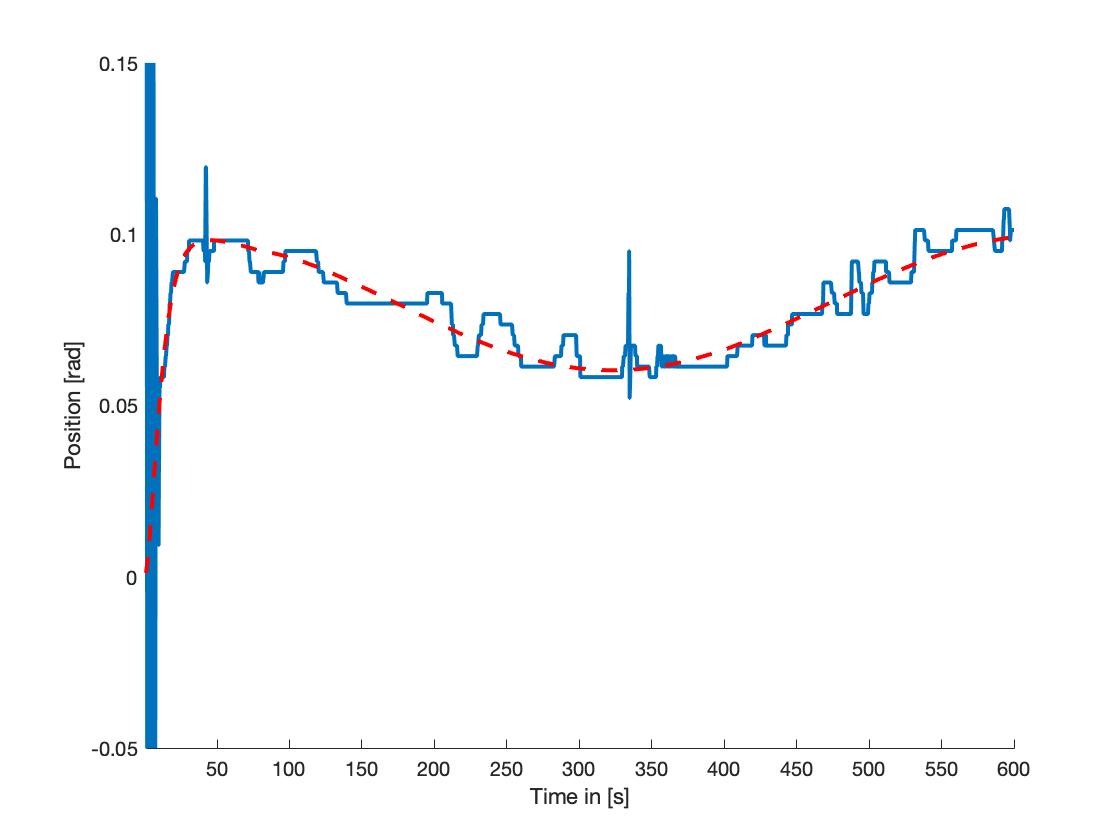,width=1\textwidth}}
\subfigure[\footnotesize Control $u_2$]
{\epsfig{figure=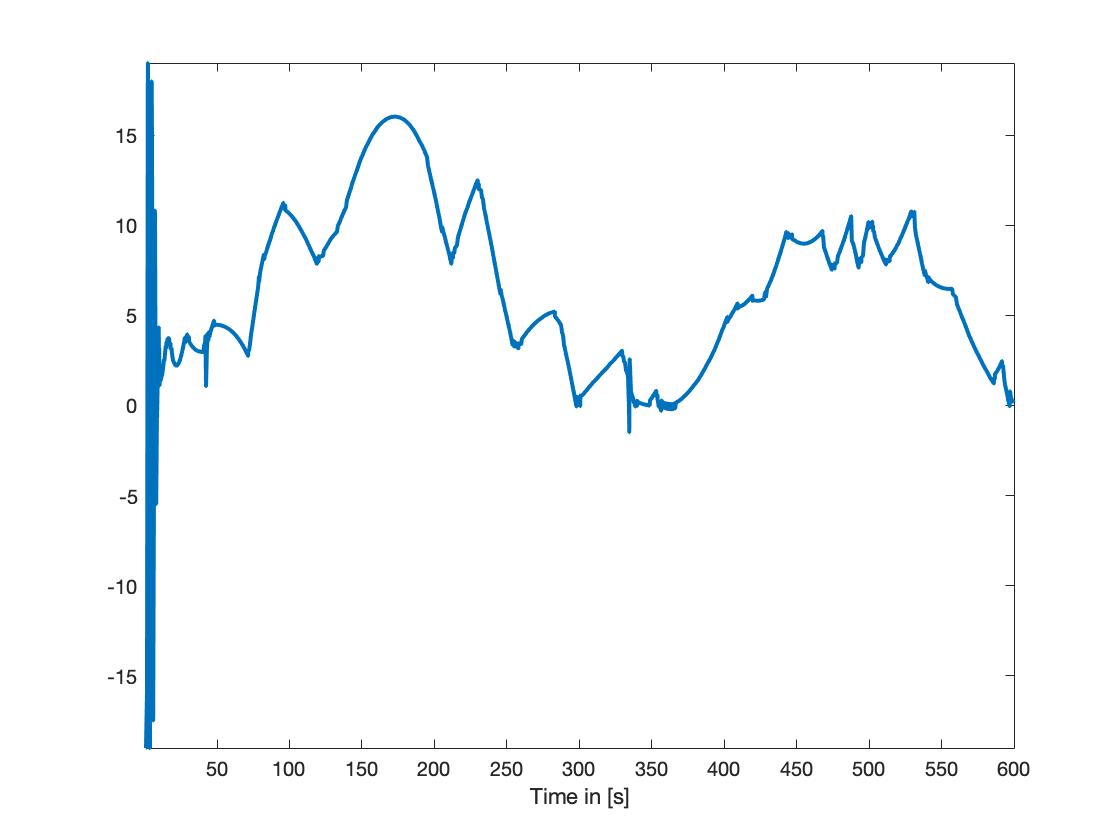,width=1\textwidth}}
\caption{Scenario 2: Pitch}\label{S22}
\end{figure*}

Consider two scenarios: 
\begin{itemize}
\item scenario 1 -- simple reference trajectory (see Fig. \ref{S11} and \ref{S12} ),
\item  scenario 2 -- complex reference trajectory (see Fig. \ref{S21} and \ref{S22}).
\end{itemize}
The tracking is excellent in both cases in spite of the rotating blades, the gyroscopic effects, and the frictions, which are all taken into account by $F_i$.

\subsubsection{Adding a mass}
\begin{figure}[!ht]
\centering%
\includegraphics[width=0.90\textwidth]{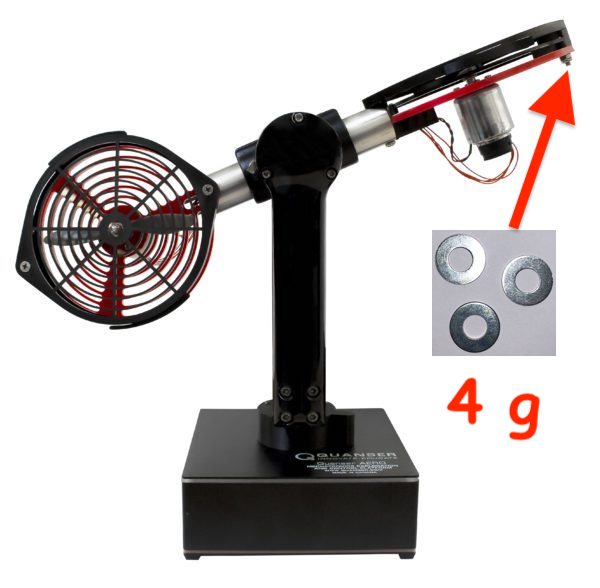}
\caption{Additive masses on the AERO}\label{maquetteM}
\end{figure}
\begin{figure*}[!b]
\centering%
\subfigure[\footnotesize Azimuth velocity (blue $--$), reference trajectory (red $- -$)]
{\epsfig{figure=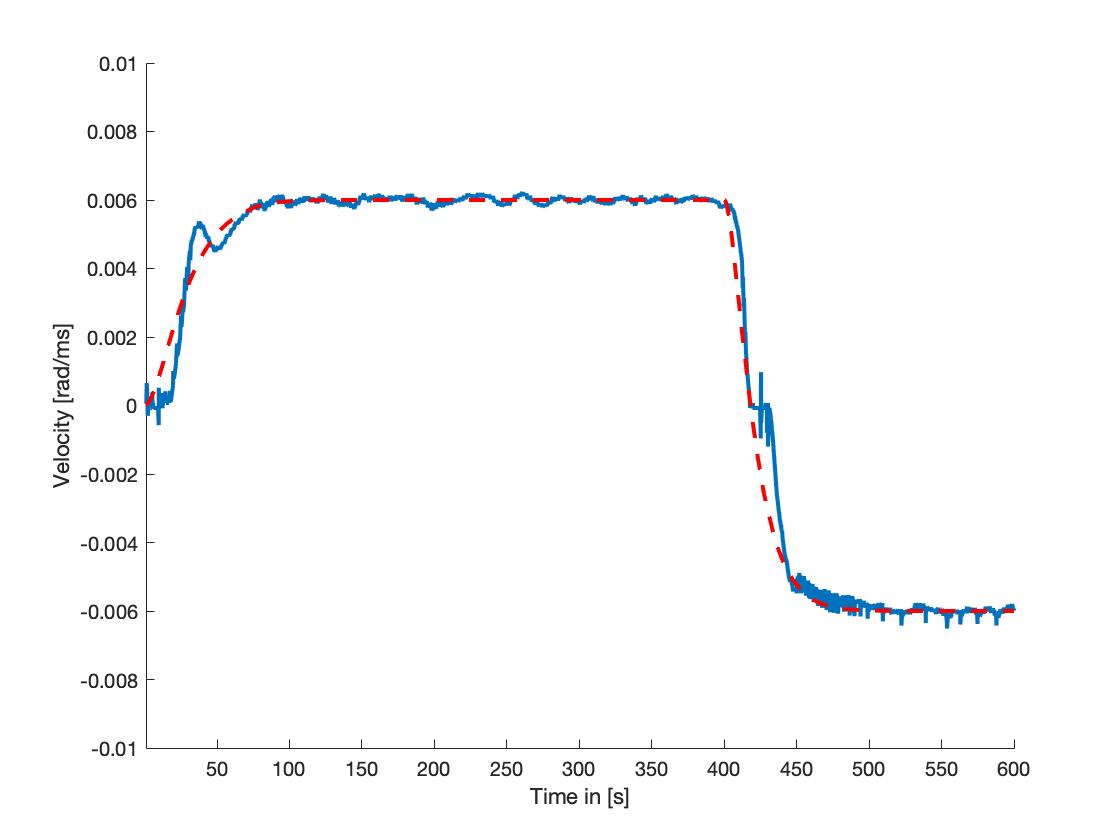,width=1\textwidth}}
\subfigure[\footnotesize Control $u_1$]
{\epsfig{figure=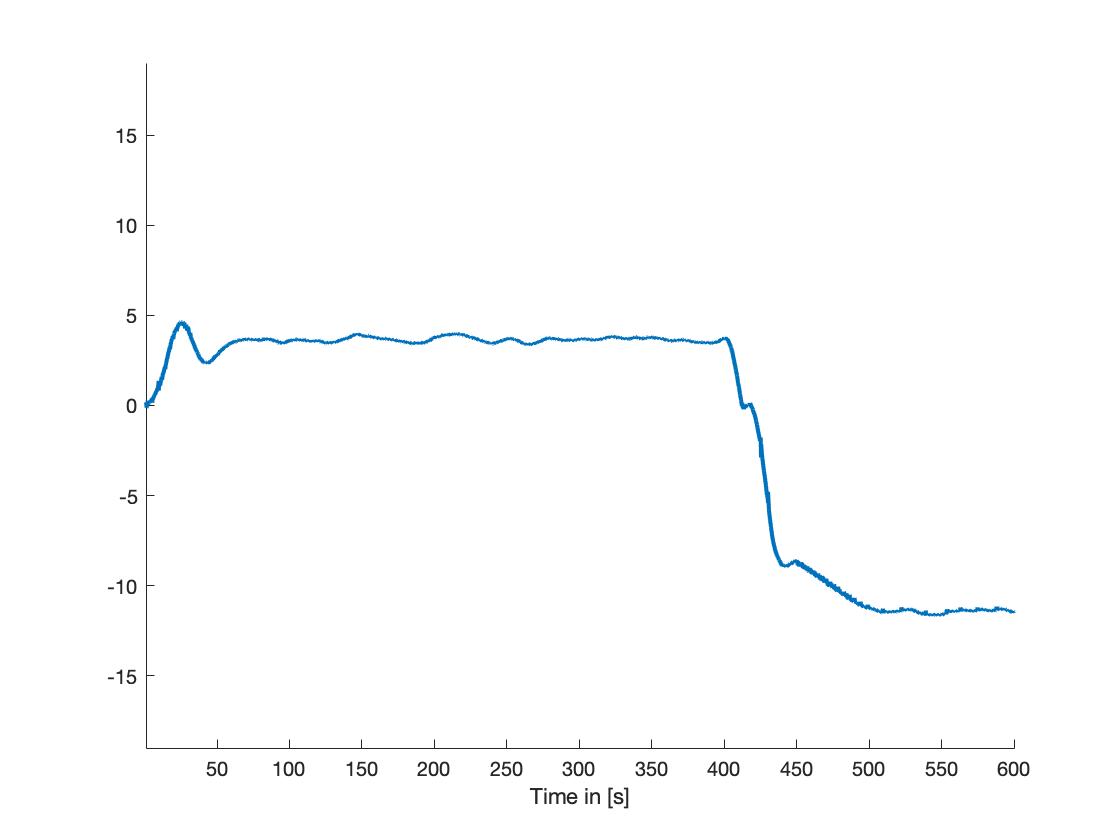,width=1\textwidth}}
\caption{Scenario 3: Azimuth}\label{S31}
\end{figure*}
\begin{figure*}[!b]
\centering%
\subfigure[\footnotesize Pitch position (blue $--$), reference trajectory (red $- -$)  ]
{\epsfig{figure=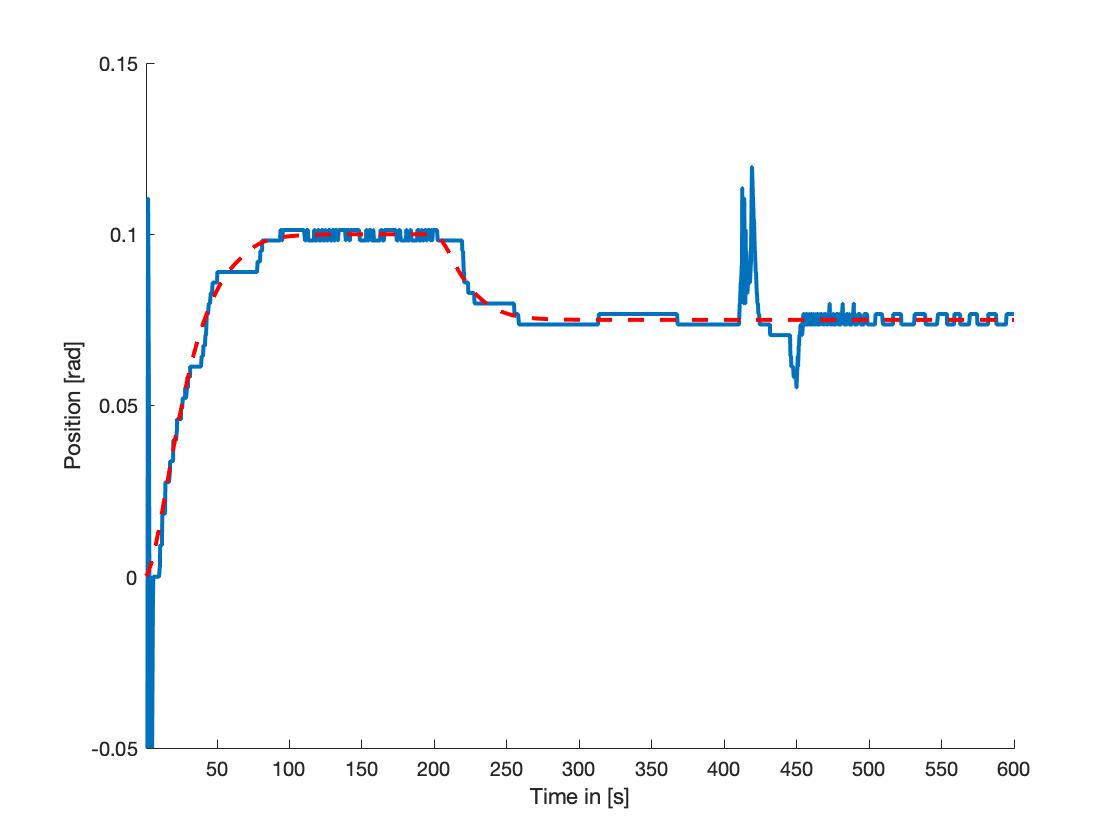,width=1\textwidth}}
\subfigure[\footnotesize Control $u_2$]
{\epsfig{figure=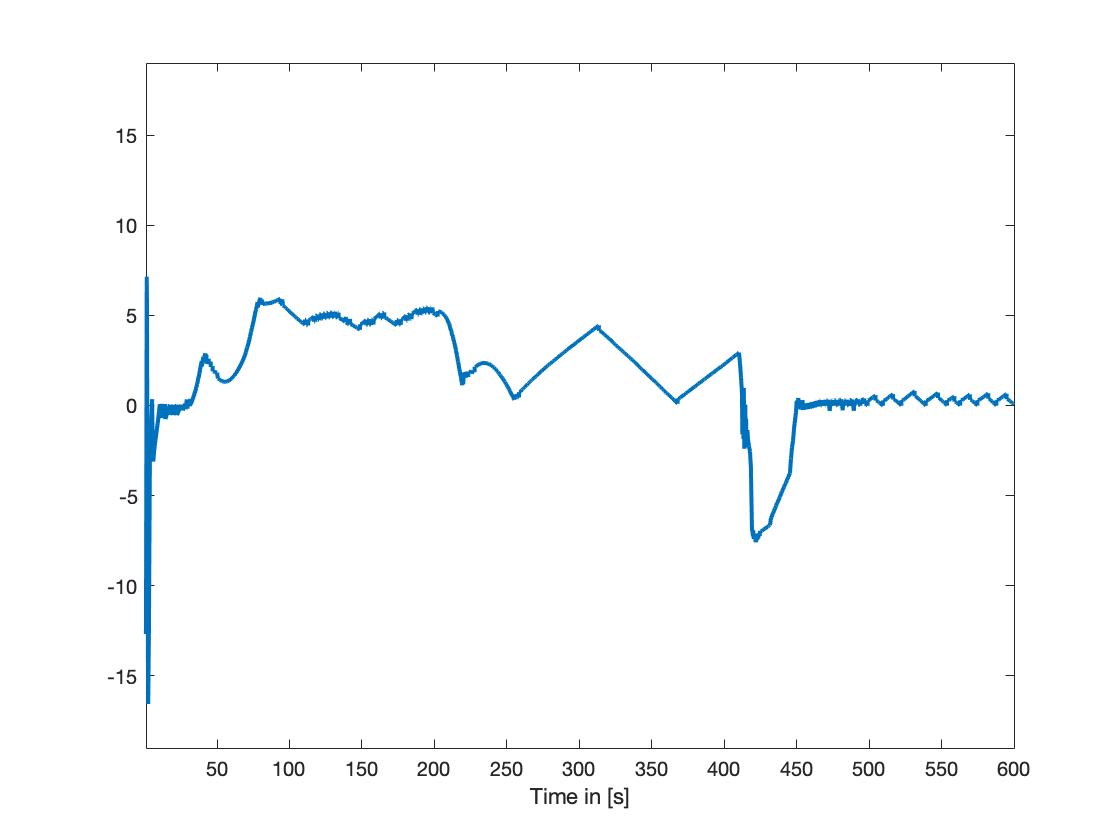,width=1\textwidth}}
\caption{Scenario 3: Pitch}\label{S32}
\end{figure*}
\begin{figure*}[!b]
\centering%
\subfigure[\footnotesize Azimuth velocity (blue $--$), reference trajectory (red $- -$)]
{\epsfig{figure=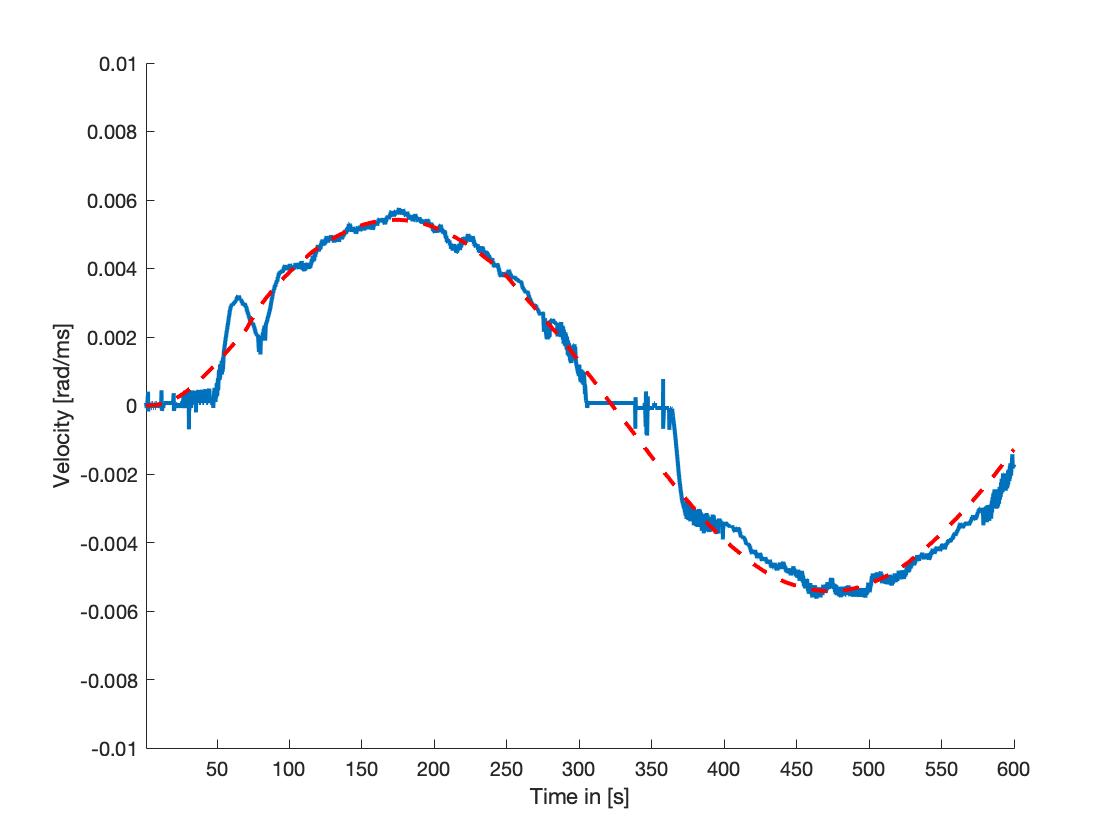,width=1\textwidth}}
\subfigure[\footnotesize Control $u_1$]
{\epsfig{figure=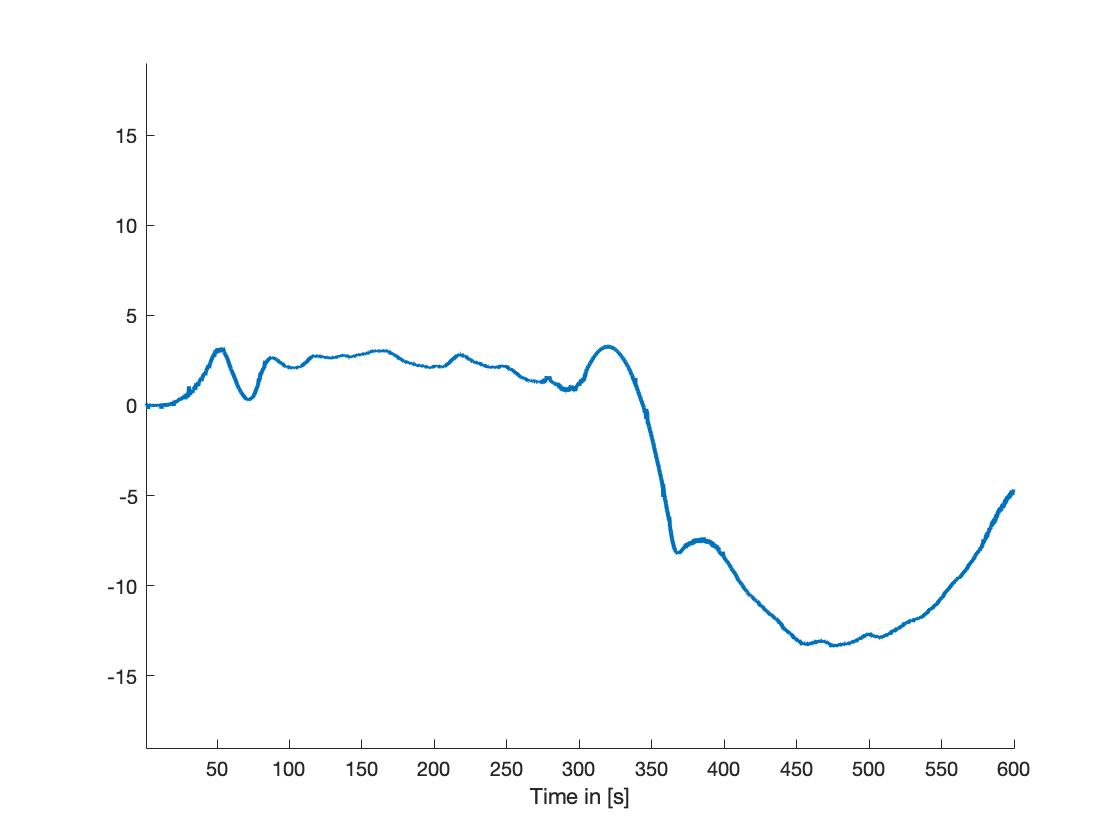,width=1\textwidth}}
\caption{Scenario 4: Azimuth}\label{S41}
\end{figure*}
\begin{figure*}[!b]
\centering%
\subfigure[\footnotesize Pitch position (blue $--$), reference trajectory (red $- -$)  ]
{\epsfig{figure=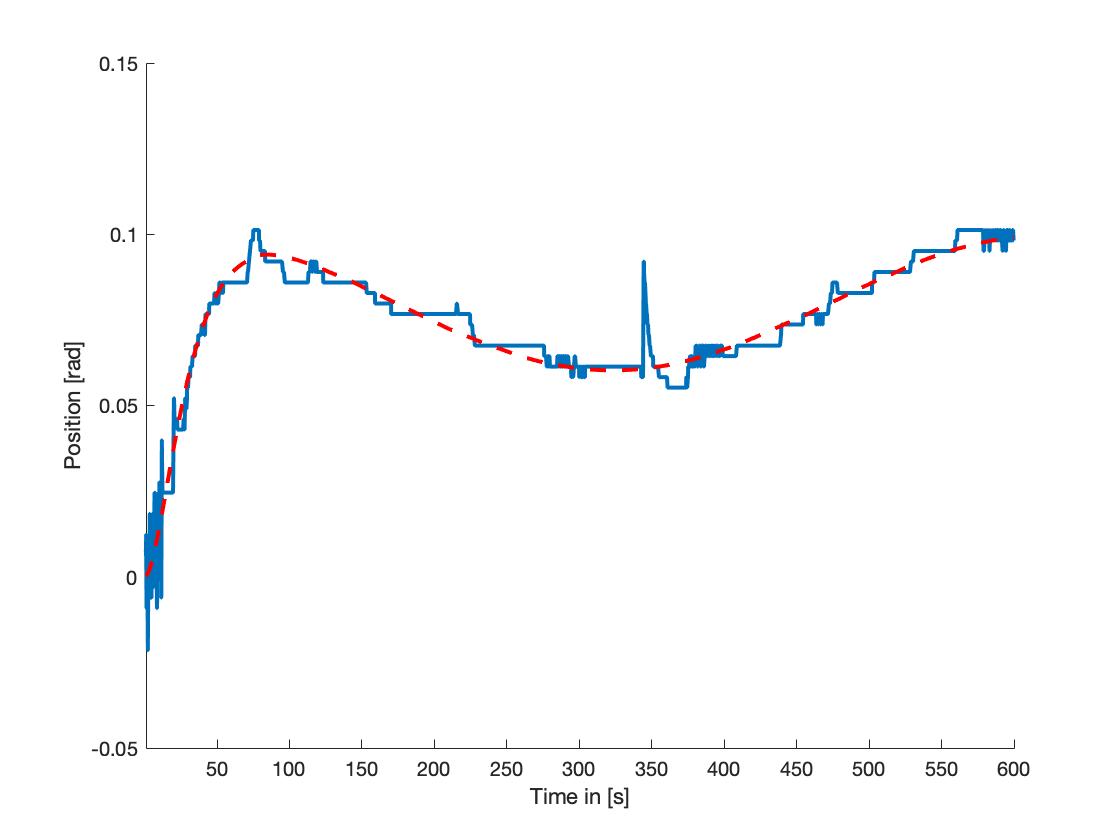,width=1\textwidth}}
\subfigure[\footnotesize Control $u_2$]
{\epsfig{figure=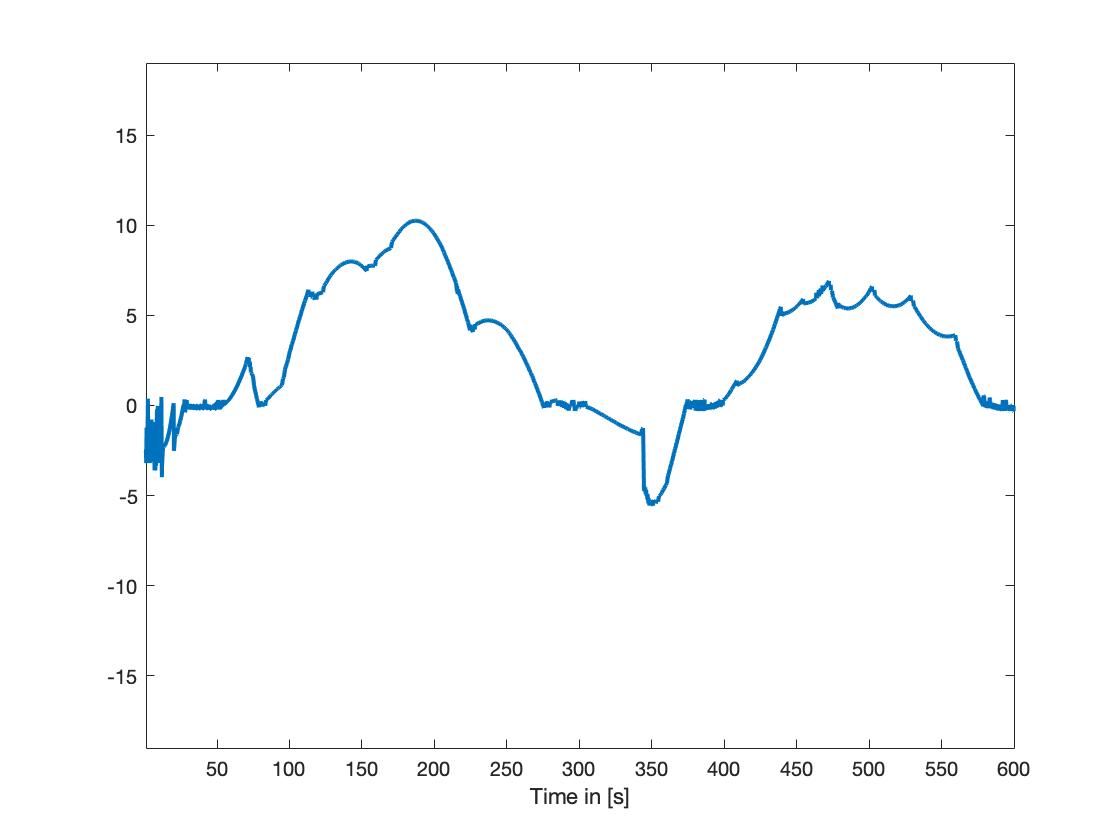,width=1\textwidth}}
\caption{Scenario 4: Pitch}\label{S42}
\end{figure*}

Fig. \ref{maquetteM}. shows that a mass of $4$ grams is added. 
It is taken into account by $F_i$, $i = 1, 2$. There is no new calibration. Keep the previous scenarios:
\begin{itemize}
\item scenario 3 -- simple reference trajectory (see Fig. \ref{S31} and \ref{S32}),
\item  scenario 4 -- complex reference trajectory (see Fig. \ref{S41} and \ref{S42}).
\end{itemize}
The tracking does not deteriorate.



\section{Conclusion}\label{conclusion}
 Of course further studies are needed in order to support the thesis of this paper. MFC might not be able to provide a satisfactory fault detection.\footnote{See \cite{csm}, \cite{toulon} for fault accommodation.} The r\^{o}le of ANNs and RL might then be compelling (see, \textit{e.g.}, \cite{lv}). 


The epistemological connections of MFC with other existing approaches in AI (see, \textit{e.g.}, \cite{russel}),  Wiener's {\em cybernetics} and {\em expert systems} for instance, will be analyzed elsewhere.


\newpage

\begin{thebibliography}{6}

\bibitem{abb}
Abbaker, A.M.O., Wang, H., Tian, Y., 2020.
\newblock Voltage control of solid oxide fuel cell power plant based on intelligent proportional integral-adaptive sliding mode control with anti-windup compensator.
\newblock Trans. Inst. Measur. Contr., 42, 116-130.

\bibitem{abou}
Aboua\"{i}ssa, H., Chouraqui, S., 2019.
\newblock On the control of robot manipulator: A model-free approach.
\newblock  J. Comput. Sci., 31, 6-16.

\bibitem{anderson}
Anderson, C.W., D.C. Hittle, D.C., Katz, A.D., Kretchmar, R.M., 1997.
\newblock Synthesis of reinforcement learning, neural networks and PI control applied to a simulated heating coil. 
\newblock Artif. Intell. Engin.,  11, 421-429.



\bibitem{astrom}
{\AA}str\"om, K.J., H\"agglund, T., 2006. 
\newblock Advanced PID Control.
\newblock Instrum. Soc. Amer..

\bibitem{murray}
{\AA}str\"om, K.J., Murray, R.M., 2008. Feedback Systems: An Introduction for Scientists and Engineers. Princeton University Press.


\bibitem{bara}
Bara, O., Fliess, M., Join, C., Day, J., Djouadi, S.M., 2018. 
\newblock  Toward a model-free feedback control synthesis for treating acute inflammation. 
\newblock J. Theoret. Biology, 448, 26-37.

 \bibitem{barth1}
Barth, J.M.O., Condomines, J.-P., Bronz, M., Hattenberger, G., Moschetta, J.-M., Join, C.,
 Fliess, M., 2020. 
 \newblock Towards a unified model-free control architecture for tail sitter micro air vehicles: Flight simulation analysis and experimental flights.
 \newblock AIAA Scitech Forum, Orlando.
 
 \bibitem{barth2}
Barth, J.M.O., Condomines, J.-P., Bronz, M., Moschetta, J.-M., Join, C., Fliess, M., 2020.
\newblock Model-free control algorithms for micro air vehicles with transitioning flight capabilities.
\newblock Int. J. Micro Air Vehic., 12. \newline {\tt https://doi.org/10.1177/1756829320914264}

\bibitem{baumeister}
Baumeister, T., Brunton, S.L., Kutz, J.N., 2018.
\newblock Deep learning and model predictive control for self-tuning mode-locked lasers. 
\newblock J. Opt. Soc. Am. B, 35, 617-626.

\bibitem{bekcheva}
Bekcheva, M., Fliess, M., Join, C., Moradi, A., Mounier, H., 2018.
\newblock Meilleure \'{e}lasticit\'{e} ``nuagique'' par commande sans mod\`{e}le.
\newblock ISTE OpenSci., 2, 15 pages. \newline  {\tt https://hal.archives-ouvertes.fr/hal-01884806/en/}

\bibitem{beltran}
Beltran-Carbajal, F., Silva-Navarro, G., Trujillo-Franco, L.G., 2018.
On-line parametric estimation of damped multiple frequency.
\newblock Elec. Power Syst. Res., 154, 423-452.


\bibitem{bourbaki}
Bourbaki, N., 1976. 
\newblock Fonctions d'une variable r\'{e}elle.
\newblock Hermann.
\newblock English translation, 2004: Functions of a Real Variable. Springer.



\bibitem{brunton}
Brunton, S.L., Noack, B.R., Koumoutsakos, P., 2020.
\newblock Machine learning for fluid mechanics.
\newblock Annu. Rev. Fluid Mech., 52, 477-508.


\bibitem{bucci}
Bucci, M.A., Semeraro, O., Allauzen, A., Wisniewski, G., Cordier, L., Mathelin, L., 2019.
\newblock Control of chaotic systems by deep reinforcement learning.
 \newblock Proc. Roy. Soc. A, 475, 20190351.

\bibitem{bu}
Bu\c{s}oniu, L., de Bruin, T., Toli\'{c}, D., Koberb, J., Palunko, I., 2018.
\newblock Reinforcement learning for control: Performance, stability, and deep approximators.
\newblock Annual Rev. Contr., 46, 8-28.

\bibitem{chen}
Chen, J., Huang, T.-C., 2004.
\newblock Applying neural networks to on-line updated PID controllers for nonlinear process control.
\newblock  J. Process Contr., 14, 211-230.

\bibitem{cheon}
Cheon, K., Kim, J., Hamadache, M., Lee, D., 2015.
\newblock On replacing PID controller with deep learning controller for DC motor system.
\newblock  J. Automat. Contr. Engin., 3, 452-456.

\bibitem{clouatre1}
Clouatre, M., Thitsa, M., (2020). 
\newblock Shaping 800nm pulses of {Y}b/{T}m co-doped laser: A control theoretic approach.
\newblock Ceramics Int., \newline {\tt https://doi.org/10.1016/j.ceramint.2020.03.123}

\bibitem{clouatre2}
Clouatre, M., Thitsa, M., (2020). 
\newblock Data-driven sliding mode control for pulses of fluorescence in STED microscopy based on F\"{o}rster resonance energy transfer pairs. 
\newblock MRS Advances, {\tt https://doi.org/10.1557/adv.2020.11}

\bibitem{clouatre3}
Clouatre, M., Thitsa, M., Fliess, M., Join, C., 2020.
\newblock A robust but easily implementable remote control for quadrotors: Experimental acrobatic flight tests.
\newblock \emph{Submitted}.


\bibitem{dierks}
Dierks, T., Jagannathan, S., 2010.
\newblock Neural network output feedback control of robot formations.
\newblock IEEE Trans. Syst. Man Cybern., 40, 383-399. 


\bibitem{duriez}
Duriez, T., Brunton, S.L., Noack, B.R., 2017. 
\newblock Machine Learning Control -- Taming Nonlinear Dynamics and Turbulence.
\newblock Springer.






 
\bibitem{bruit}
Fliess, M., 2006. 
\newblock Analyse non standard du bruit.
\newblock  C.R. Acad. Sci. Paris Ser. I, 342, 797-802.

\bibitem{intel}
Fliess, M., Join, C., 2008. 
\newblock Intelligent PID controllers.
\newblock 16th Med. Conf. Contr. Automat., Ajaccio. {\tt https://hal.inria.fr/inria-00273279/en/}


\bibitem{csm}
Fliess, M., Join, C., 2013. Model-free control.
Int. J. Contr., 86, 2228-2252.





\bibitem{sira1}
Fliess, M., Sira-Ram\'{\i}rez, H., 2003.
\newblock An algebraic framework for linear identification.
\newblock ESAIM Contr. Optimiz. Calc. Variat.,
9, 151-168.

\bibitem{sira2}
Fliess, M., Sira-Ram\'{\i}rez, H., 2008.
\newblock Closed-loop parametric identification for
  continuous-time linear systems via new algebraic techniques.
Garnier, H., Wang, L. (Eds): Identification of
  Continuous-time Models from Sampled Data, Springer,
 pp. 362-391. 

\bibitem{had}
Haddar, M., Chaari, R.,  Baslamisli, S.C., Chaari, F., Haddar, M., 2019.
\newblock Intelligent PD controller design for active suspension system based on robust model-free control strategy.
\newblock J. Mech. Engin. Sci., 233, 4863-4880.

\bibitem{han}
Han, S., Wang, H., Tian, Y., 2020.
\newblock A linear discrete-time extended state observer-based intelligent PD controller for a 12 DOFs lower limb exoskeleton LLE-RePA.
\newblock Mech. Syst. Sign. Proc., 138, 106547.

\bibitem{hat}
Hatipoglu, K., Olama, H., Xue, Y., 2020.
\newblock Model-free dynamic voltage control of a synchronous generator-based microgrid.
\newblock  IEEE Innov. Smart Grid Techno. Conf., Washington.

\bibitem{hong}
Hong, Y., Yang, W., Jiang, B., Yan, X.-G., 2020.
\newblock A novel multi-agent model-free control for state-of-charge balancing between distributed battery energy storage systems.  
\newblock IEEE Trans. Emerg. Topics Comput. Intel., {\tt doi: 10.1109/TETCI.2020.2978434}
 
\bibitem{hwang}
Hwangbo, J., Sa, I., Siegwart, R., Hutter, M., 2017. 
\newblock Control of a quadrotor with reinforcement learning. 
\newblock IEEE Robot. Automat. Lett., 2, 2096-2103. 






\bibitem{hardware}
Join, C., Chaxel, F., Fliess, M., 2013.
\newblock ``Intelligent'' controllers on cheap and small programmable devices.
\newblock 2nd Int. Conf. Contr. Fault-Tolerant Syst., Nice. \newline
{\tt https://hal.archives-ouvertes.fr/hal-00845795/en/}

\bibitem{iot}
Join, C., Fliess, M,  Chaxel, F., 2020.
\newblock Model-Free Control as a Service in the Industrial Internet of Things: Packet loss and latency issues via preliminary experiments.
\newblock 16th Med. Conf. Contr. Automat., Saint-Rapha\"{e}l. \newline
{\tt https://hal.archives-ouvertes.fr/hal-02546750/en/}





\bibitem{kahn}
Kahn, S.G., Hermann, G., Lewis, F.L., Pipe, T., Melhuish, C., 2012. 
\newblock Reinforcement learning and optimal adaptive control: An overview and implementation examples.
\newblock Annu. Rev. Contr., 36, 42-52.

\bibitem{kizir}
Kizir, S., Bing\"{u}l, Z., 2019.
\newblock Design and development of a Steward platform assisted and navigated transphenoidal surgery.
\newblock Turk. J. Elec. Eng. Comp. Sci., 27, 961-972.


\bibitem{kolmogorov}
Kolmogorov, A.N., Fomin, S.V., 1957 \& 1961. 
\newblock Elements of the Theory of Functions and Functional Analysis, vol. 1 \& 2 (translated from the Russian).
\newblock Graylock.


\bibitem{kiumarsi}
Kiumarsi, B., Vamvoudakis, K.G., Modares, H., Lewis, F.L., 2018.
\newblock Optimal and autonomous control using reinforcement learning: A survey.
\newblock IEEE Trans. Neural Netw. Learn. Syst., 29, 2042-2062.



\bibitem{toulon}
Lafont, F., Balmat, J.-F., Pessel, N., Fliess, M., 2015.
\newblock A model-free control strategy for an experimental greenhouse with an application to fault accommodation. 
Comput. Electron.  Agricul., 110, 139-149.

\bibitem{lambert}
Lambert, N.O., Drew, D.S., Yaconelli, J., Levine, S., Calandra, R., Pister, K.S.J., 2019.
\newblock Low-level control of a quadrotor with deep model-based reinforcement learning. 
\newblock IEEE Robot. Automat. Lett., 4, 4224-4230.


\bibitem{lecun1}
Le Cun, Y., 2019.
\newblock Quand la machine apprend.
\newblock Odile Jacob, 2019.

\bibitem{lecun2}
LeCun, Y., Bengio, Y., Hinton, G., 2015.
\newblock Deep learning.
\newblock Nature, 521, 436-444.

\bibitem{li}
Li, S., Zhang, Y., 2018.
\newblock  Neural Networks for Cooperative Control of Multiple Robot Arms.
\newblock  Springer.

\bibitem{lucia}
Lucia, S., Karg, B., 2018.
\newblock  A deep learning-based approach to robust nonlinear model predictive control.
IFAC PapersOnLine, 51-20, 511-516.

\bibitem{luo}
Luo, B., Liu, D., Huang, T., Wang, D., 2016. 
\newblock Model-free optimal tracking control via critic-only Q-learning. 
\newblock IEEE Trans. Neural Netw. Learn. Syst., 27, 2134-2144.

\bibitem{lv}
Lv, F., Wen, C., Bao, Z., Liu, M., 2016.
\newblock Fault diagnosis based on deep learning.
\newblock \emph{Amer. Contr. Conf.}, Boston.

\bibitem{ma}
N. Ma, G. Song, H.-J. Lee.
\newblock Position control of shape memory alloy actuators with internal electrical resistance feedback using neural networks.
\newblock Smart Mater. Struct., 13, 777-783, 2004.



\bibitem{matni1}
Matni, N., Proutiere, A., Rantzer, A., Tu, S., 2019.
\newblock From self-tuning regulators to reinforcement learning and back again.
\newblock 58th Conf. Decis. Contr., Nice, 2019.

\bibitem{matni2}
Matni, N., Tu, S., 2019.
\newblock A tutorial on concentration bounds for system identification.
\newblock 58th Conf. Decis. Contr., Nice.

\bibitem{menhour}
Menhour L., d'Andr\'{e}a-Novel, B., Fliess, M., Gruyer, D., Mounier, H., 2018.
\newblock An efficient model-free setting for longitudinal and lateral vehicle control: Validation through
the interconnected Pro-SiVIC/RTMaps.
\newblock IEEE Trans. Intel. Transp. Syst., {19}, 461-475.

\bibitem{mich}
Michailidis, I.T., Schild, T., Sangi, R., Michailidis, P., Korkas, C., F\"{u}tterer, J., M\"{u}ller, D., Kosmatopoulos, E.B., 2018.
\newblock Energy-efficient HVAC management using cooperative, self-trained, control agents: A real-life German building case study. 
\newblock App. Ener., 211, 113-125.


\bibitem{miller}
Miller III, W.T.,, Sutton, R.S., Werbos, P.J., 1990 (Eds).
\newblock Neural Networks for Control.
\newblock MIT Press.

\bibitem{mnih}
Mnih, V., Kavukcuoglu, K., Silver, D., Rusu, A.A., Veness, J., Bellemare, M.G., Graves, A., Riedmiller, M.,  Fidjeland, A.K., Ostrovski, G., Petersen, S., Beattie, C., Sadik, A., Antonoglou, I., King, H., Kumaran, D., Wierstra, D., Legg, S., Hassabis, D., 2015. 
\newblock Human-level control through deep reinforcement learning. 
\newblock Nature, 518, 529-533.

\bibitem{moe}
Moe, S., Rustand, A.M., Hanssen, K.G., 2018.
\newblock Machine learning in control systems: An overview of the state of the art.
\newblock M. Bramer, M. Petridis (Eds): Artificial Intelligence XXXV, Lect. Notes Artif. Intel. 11311, pp. 250-264, Springer.


\bibitem{nd}
N'Doye, I., Asiri, S., Aloufi, A., Al-Awan, A., Laleg-Kirati, T.-M., 2020. 
\newblock Intelligent proportional-integral-derivative control-based modulating functions for laser beam pointing and stabilization, 
\newblock IEEE Trans. Contr. Syst. Techno., 28,1001-1008.

\bibitem{nicol}
Nicol, C., Macnab, C.J.B., Ramirez-Serrano, A., 2008.
\newblock Robust neural network control of a quadrotor helicopter.
\newblock Canad. Conf. Elec. Comput. Engin., Niagara Falls, 2008.

\bibitem{plumejeau}
Plumejeau, B., Delprat, S., Keirsbulck, L., Lippert, M., Abassi, W., 2019.
\newblock Ultra-local model-based control of the square-back Ahmed body wake flow.
\newblock Phys. Fluids, 31, 085103.

\bibitem{qin}
Qin, Z.-C., Xin, Y., Sun, J.-Q., 2020.
\newblock Dual-loop robust attitude control for an aerodynamic system with unknown dynamic model: algorithm and experimental validation.
\newblock IEEE Access, 8, 36582-36594.

\bibitem{qu}
Qu, S.T., 2019. 
\newblock Unmanned powered paraglider flight path control based on PID neutral network.
\newblock IOP Conf. Ser. Mater. Sci. Eng., 470, 012008.

\bibitem{rabault}
Rabault, J., Kuchta, M., Jensen, A., R\'eglade, U., Cerardi, N., 2019.
\newblock Artificial neural networks trained through deep reinforcement learning discover control strategies for active flow control. 
\newblock J. Fluid Mech., 865, 281-302.









\bibitem{radac}
Radac, M.-B., Precup, R.-E., Roman, R.-C., 2017.
\newblock Model-free control performance improvement using virtual reference feedback tuning and reinforcement Q-learning.
\newblock Int. J. Syst. Sci., 48, 1071-1083. 


\bibitem{rampazzo}
Rampazzo, M., Tognin, D., Pagan, M., Carniello, L., Beghi, A., 2019.
\newblock Modelling, simulation and real-time control of a laboratory tide generation 
\newblock Contr. Eng. Pract., 83, 165-175.



\bibitem{recht}
Recht, B., 2019. 
\newblock A tour of reinforcement learning: The view from continuous control.
\newblock Annu. Rev. Contr. Robot. Autonom. Syst., 2, 253-279.

\bibitem{rocher}
Rocher, V., Join, C., Mottelet, S., Bernier, J., Rechdaoui-Guerin, S., Azimi, S., Lessard, P., Pauss, A., Fliess, M., 2018.
\newblock La production de nitrites lors de la d\'{e}nitrification des eaux us\'{e}es par biofiltration - strat\'{e}gie de contr\^{o}le et de r\'{e}duction des concentrations r\'{e}siduelles.
\newblock J. Water Sci., 31, 61-73.



 \bibitem{russel}
Russel, S., Norvig, P., 2016.
\newblock Artificial Intelligence -- A Modern Approach (3rd ed.).
\newblock Pearson.

\bibitem{san}
Sancak, C., Yamac, F., Itik, M., Alici, G., 2019.
\newblock  Model-free control of an electro-active polymer actuator.
\newblock Mater. Res. Expr., 6, 055309.

\bibitem{sej}
Sejnowski, T.J., 2020.
\newblock The unreasonable effectiveness of deep learning in artificial intelligence.
\newblock  Proc. Nat. Acad. Sci., {\tt https://doi.org/10.1073/pnas.1907373117}


\bibitem{go}
Silver, D., Huang, A., Maddison, C.J., Guez, A., Sifre, L., van den Driessche, G., Schrittwieser, J., Antonoglou, I., Panneershelvam, V., Lanctot, M., Dieleman, S., Grewe, D., Nham, J., Kalchbrenner, N., Sutskever, I., Lillicrap, T., Leach, M., Kavukcuoglu, K., Graepel, T., Hassabis, D., 2016.
\newblock Mastering the game of Go with deep neural networks and tree search.
\newblock Nature, 529, 484-489. 


\bibitem{sira}
Sira-Ram\'{\i}rez, H., Garc\'{\i}a-Rodr\'{\i}guez, C., Cort\`{e}s-Romero, J., Luviano-Ju\'{a}rez, A., 2014.  Algebraic Identification and Estimation Methods in Feedback Control Systems. Wiley.

 
 \bibitem{stalph}
Stalph, P., 2014.
\newblock Analysis and Design of Machine Learning Techniques
\newblock Springer.
 
 

\bibitem{sugiyama}
Sugiyama, M., 2015. 
\newblock  Statistical Reinforcement Learning -- Modern Machine Learning Approaches
\newblock CRC Press.


\bibitem{sutton}
Sutton, R.S., Barto, A.G., 2018.
\newblock Reinforcement Learning (2nd ed.).
\newblock MIT Press.


\bibitem{tich}
Ticherfatine, M., Zhu, Q., 2018.
\newblock Fast ferry smoothing motion via intelligent PD controller. 
\newblock \emph{J. Marine. Sci. App.}, 17, 273-279.

\bibitem{villagra}
Villagra, J., Join, C., Haber, R., Fliess, M., 2020.
\newblock Model-free control for machine tool systems.
\newblock 21st World IFAC, Berlin. \newline {\tt https://hal.archives-ouvertes.fr/hal-02568336/en/}




\bibitem{wang0}
Wang, Y., Li, H., Liu, R., Yang, L., Wang, X., 2020.
\newblock Modulated model-free predictive control with minimum switching losses for PMSM
drive system.
\newblock IEEE Access, 8, 20942-20953.

\bibitem{wang1}
Wang, H., Li, S., Tian, Y., Aitouche, A., 2016.
\newblock Intelligent proportional differential neural network control for unknown nonlinear system.
\newblock Stud. Informat. Contr., 25, 445-452.

\bibitem{wang2}
Wang, Y., Velswamy, K., Huang, B., 2018.
\newblock A novel approach to feedback control via deep reinforcement learning.
\newblock IFAC PapersOnLine, 51-18, 31-36.

\bibitem{wang3}
Wang, Z., Wang, J., 2020. 
\newblock Ultra-local model predictive control: A model-free approach and its application on automated vehicle trajectory tracking.
\newblock Contr. Eng. Pract., 101, 104482.

\bibitem{weislander}
Waslander, S.L., Hoffmann, G.M., Jung Soon Jang ; Tomlin, C.J., 2005.
\newblock Multi-agent quadrotor testbed control design: integral sliding mode vs. reinforcement learning.
\newblock  IEEE/RSJ Int. Conf. Intell. Robot. Syst., Edmonton.

\bibitem{wu}
 Wu, Y., Song, Q., Yang, X., 2007.
\newblock Robust recurrent neural network control of biped robot.
\newblock J. Intell. Robot. Syst., 49, 151-169.

\bibitem{yang}
Yang, H., Liu, C., Shi, J., Zhong, G., 2019.
\newblock Development and control of four-wheel independent driving and modular steering electric vehicles for improved maneuverability limits.
\newblock SAE Tech. Paper, 2019-01-0459.

\bibitem{yosida}
Yosida, K., 1984. Operational Calculus (translated from the
Japanese). Springer.

\bibitem{zhang1}
Zhang, Y., Ding, S.X., Yang, Y., Li, L., 2015.
\newblock  Data-driven design of two-degree-of-freedom controllers using reinforcement learning techniques.
\newblock  IET Contr. Theory Appli., 9,  1011-1021.

\bibitem{zhang1bis}
Zhang, J., Jin, J., Huang, L., 2020.
\newblock Model-free predictive current control of PMSM drives based on extended state observer using ultra-local model.
\newblock IEEE Trans. Indus. Electron., {\tt doi: 10.1109/TIE.2020.2970660}

\bibitem{zhang2}
Zhang, X., Li, M., Ding, H., Yao, X., 2019.
\newblock Data-driven tuning of feedforward controller structured with infinite impulse response filter via iterative learning control.
\newblock  IET Contr. Theory Appli., 13, 1062-1070.

\bibitem{zhang2bis}
Zhang, Y., Liu, X., Liu, J., Rodriguez, J., Garcia, C., 2020.
\newblock Model-free predictive current control of power converters based on ultra-local model.
\newblock IEEE Int. Conf. Indust.Techno., Buenos Aires.

\bibitem{zhang3}
Zhang, X., Wang, H., Tian, Y., Peyrodie, L., Wang, X., 2018.
\newblock Model-free based neural network control with time-delay estimation for lower extremity exoskeleton.
\newblock Neurocomput., 272, 178-188.


\bibitem{zhang4}
Zhang, X.-M., Wei, Z.,, Asad, R., Yang, X.-C., Wang, X., 2019.
\newblock When does reinforcement learning stand out in in control? A comparative study on state representation.
\newblock npj Quantum Inf., 5.  {\tt https://doi.org/10.1038/s41534-019-0201-8}

\bibitem{zhu}
Zhu, L., Ma, J., Wang, S., 2019.
\newblock Deep neural networks based real-time optimal control for lunar landing.
\newblock IOP Conf. Ser. Mater. Sci. Engin., 608, 012045.


\end{thebibliography}
\end{document}